\documentclass[twocolumn,showpacs,preprintnumbers,amsmath,amssymb]{revtex4}

\usepackage{graphicx}
\usepackage{dcolumn}
\usepackage{bm}

\begin{document}

\title{
Quantum melting of charge ice and non-Fermi-liquid behavior: \\
An exact solution for the extended Falicov-Kimball model in the ice-rule limit
}

\author{Masafumi Udagawa}
\author{Hiroaki Ishizuka}
\author{Yukitoshi Motome}%
\affiliation{%
Department of Applied Physics, 
University of Tokyo, 
Tokyo 113-8656, Japan}%

\date{\today}

\begin{abstract}
An exact solution is obtained for 
a model of itinerant electrons coupled to ice-rule variables  
on the tetrahedron Husimi cactus, 
an analogue of the Bethe lattice of corner-sharing tetrahedra. 
It reveals 
a quantum critical point 
with the emergence of non-Fermi-liquid behavior 
in melting of the ``charge ice" insulator. 
The electronic structure is compared with 
the numerical results 
for the pyrochlore-lattice model 
to elucidate 
the physics of electron systems interacting with the tetrahedron ice rule. 
\end{abstract}

\pacs{71.10.Fd, 71.10.Hf, 71.20.-b, 71.23.-k}
\maketitle
The ice rule is a local constraint observed in a broad range of systems in condensed matter physics. 
It imposes a configurational constraint on two-state variables defined at neighboring four lattice sites 
so that two out of four are in the opposite state to the other two. 
The most well-known material is water ice,  
in which the two states correspond to the configuration of hydrogens~\cite{Bernal1933,Pauling}.  
An analogy was drawn by Anderson in the cation ordering of Fe$^{2+}$ and Fe$^{3+}$ in magnetite Fe$_3$O$_4$~\cite{Anderson}. 
More recently, a magnetic analogue was found in several pyrochlore oxides, the so-called spin ice, 
such as Ho$_2$Ti$_2$O$_7$~\cite{Harris} and Dy$_2$Ti$_2$O$_7$~\cite{Ramirez}. 

The ice rule enforces local correlations; however, 
it is underconstraint and not enough to make the entire system ordered. 
The ground-state manifold retains macroscopic degeneracy, resulting in residual entropy~\cite{Pauling,Ramirez}.
Nevertheless, 
the ice-rule configuration is not completely disordered but cooperative in nature:
There is 
a spatial power-law correlation in the two-state variables 
originating from a hidden gauge structure~\cite{Huse}. 
Considerable progress on the understanding of such cooperative aspects has been made in the last decade through the study of spin ice~\cite{Bramwell}. 

In contrast to such ``localized spin physics", 
much less is known for the role of the ice rule in itinerant systems. 
It is intriguing to elucidate how the cooperative nature from the ice-rule constraint
affects the electronic and transport properties.
The issue will also be experimentally relevant to a wide range of pyrochlore-based compounds, 
such as mixed-valence compounds 
with a charge-ordering tendency~\cite{Kondo1997,Takeda2005,Blaha2004} and  
itinerant $d$-electron materials including Ising-like 
rare-earth moments~\cite{Taguchi,Nakatsuji}. 
Only a few theoretical studies have been carried out so far~\cite{Chen,Fulde}.

\begin{figure}[b]
\begin{center}
\includegraphics[width=0.4\textwidth]{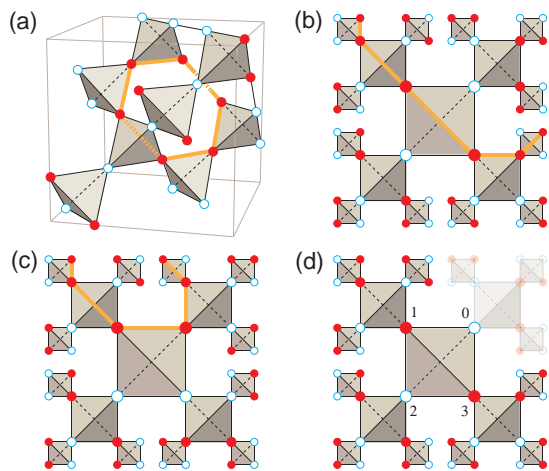}
\end{center}
\caption{\label{graph} 
(color online). A typical ice-rule configuration for (a) pyrochlore lattice and (b) tetrahedron Husimi cactus (THC).  
The sites with $n_i^f=1$ ($0$) are shown by filled (open) circles. An example of loop and one-dimensional chain 
connecting the sites with $n_i^f=1$ is shown by bold orange lines. (c) An apparently different ice-rule configuration 
obtained from (b) by interchanging the upper right and lower right branches.
(d) A branch of THC considered in the calculations of $g$ in Eq.~(\ref{perturb}).
}
\end{figure}
 
In this Letter, we address this issue in one of the simplest models 
which describe fermions interacting with ice-rule variables, an extended Falicov-Kimball model. 
We obtain an {\it exact} solution to this model on the Husimi cactus of tetrahedra, i.e., 
an analogue of the Bethe lattice composed of corner-sharing tetrahedra. 
We clarify the ground-state phase diagram including 
a ``charge ice" insulator in which the fermions are localized in the ice-rule configuration. 
The solution reveals that a non-Fermi-liquid behavior appears at 
a quantum critical point (QCP) where the charge ice melts as the interaction decreases.
By comparison with the numerical results for 
the pyrochlore lattice, we show that our Husimi cactus model  
provides a cornerstone of itinerant ice-rule problems. 

We start with the extended Falicov-Kimball model on the pyrochlore lattice [Fig.~\ref{graph}(a)], 
\begin{equation}
\mathcal{H} = 
-t\sum\limits_{\langle i,j\rangle}(c_i^{\dag}c_j + {\rm H.c.}) 
+U\sum\limits_i n^c_i (n^f_i-\frac{1}{2}) + V\sum\limits_{\langle i,j\rangle}n^f_i n^f_j, 
\label{Ham}
\end{equation}
where the first term describes the hopping of spinless fermions $c$, 
the second term represents the onsite repulsion 
between spinless fermions and immobile particles $f$, and 
the last term is the repulsion between immobile particles. 
Here, $n^c_i = c_i^{\dag}c_i$, $n^f_i = 0$ or $1$ (c number), 
and the sums $\langle i,j\rangle$ are taken over the nearest-neighbor sites. 
Hereafter, we focus on the case in which 
the immobile particles satisfy the ice rule; see Fig.~\ref{graph}(a).
This is achieved by setting $\sum_i n^f_i = N/2$ ($N$ is the total number of sites) and $V \to \infty$. 

The partition function of the model is calculated by 
$
Z = {\rm Tr}_f {\rm Tr}_c \exp(-\beta \mathcal{H})
$, 
where ${\rm Tr}_f$ (${\rm Tr}_c$) is the trace over the immobile-particle (spinless-fermion) degree of freedom, 
and $\beta = 1/k_{\rm B} T$ is the inverse temperature with the Boltzmann constant $k_{\rm B}$. 
For a given configuration of immobile particle $\{ n^f_i \}$, the Hamiltonian (\ref{Ham}) is reduced to a one-body problem given by 
\begin{eqnarray}
\mathcal{H} (\{n^f_i\}) = -t \sum\limits_{\langle i,j\rangle}
(c_i^{\dag} c_j + {\rm H.c.}) 
+ \sum\limits_i U_i n^c_i,
\label{effHam}
\end{eqnarray}
where $U_i$ is the onsite potential determined by the configuration $\{n^f_i\}$ 
as $U_i=U(n^f_i-1/2)$. Then the trace ${\rm Tr}_f$ in the partition function is replaced by the sum over $\{ n^f_i \}$ 
which satisfies the ice-rule constraint: 
$
Z = \sum_{\{n^f_i\} \, \in \, {\sf ice}} {\rm Tr}_c \exp [-\beta \mathcal{H} (\{n^f_i\}) ].
$
This is, in principle, feasible to calculate because ${\rm Tr}_c$ is performed by 
a diagonalization of the one-body $N \times N$ Hamiltonian (\ref{effHam}), but in practice, 
it is difficult for large system sizes because the sum within the ice manifold 
increases exponentially $\sim 1.5^{N/2}$~\cite{Pauling}.

A dramatic simplification is introduced 
by considering a modified structure of the pyrochlore lattice, that is, a Husimi cactus of tetrahedra~\cite{Husimi}. 
It is an analogue of the Bethe lattice composed of corner-sharing tetrahedra, as shown in Fig.~\ref{graph}(b), 
which we call here the tetrahedron Husimi cactus (THC). 
THC shares two important structural features with the pyrochlore --- 
the tetrahedral units and their corner-sharing network. 
A difference is in the global connection of tetrahedra: 
The pyrochlore lattice has loops running across different tetrahedra 
[see Fig.~\ref{graph}(a)], but THC does not have such global loops. 
Despite of this difference, theTHC model captures several essential features of the pyrochlore, as we will see later. 

The simplification by considering THC is twofold. First, while the ice-rule configurations in THC are also macroscopically degenerate ($=6\times 3^{N/3}$), 
they are topologically equivalent because one can relabel the site numbers by interchanging branches which spread from the same tetrahedron
[see Figs.~\ref{graph}(b) and (c)]. 
Consequently, all the possible ice-rule configurations give 
an identical Boltzmann weight, 
and therefore, the sum over $\{n^f_i\}$ can be suppressed in the calculations of any observable. 

On top of that, 
the second crucial point is that for any ice-rule configuration on THC 
we can obtain one-body Green's functions {\it exactly} 
by using recursion equations similar to those often used 
in the Bethe lattice problems~\cite{Thorpe,Eckstein,note-cpa}. 
To see this explicitly, let us consider the 
$T=0$ local retarded Green's function at the site $i$, $G_{i}(\varepsilon) \equiv t \langle i|
[\varepsilon - 
\mathcal{H} (\{ n^f_i \}) 
+i\delta]^{-1} |i\rangle$ 
($t$ is included to make $G$ dimensionless). 
Similarly, we define the Green's function $g_i(\varepsilon)$ for a branch given by 
terminating a half of the tetrahedral network at the site $i$, 
as shown in Fig.~\ref{graph}(d). $g_0$ is formally written by the expansion 
in terms of the hopping: 
\begin{align}
& g_0 = g^{(0)}_0 + g_0^{(0)}  
(g_{1}+g_{2}+g_{3})  
g_0\nonumber\\
&- g_0^{(0)} 
\bigl[g_{1} 
(g_{2}+g_{3}) + g_{2} 
(g_{3}+g_{1}) 
+ g_{3}  
(g_{1}+g_{2})\bigr] 
g_0 + \cdots,
\label{perturb}
\end{align}
where $g^{(0)}_0 = t(\varepsilon-U_0)^{-1}$ is 
the atomic Green's function. Here, the second term corresponds to the processes 
where a fermion hops from the site $i=0$ to one of the other three sites in the same tetrahedron, $j=1$, $2$, or $3$, then
propagates within the branch belonging to the site $j$, 
and returns to the original site $i=0$ [see Fig.~\ref{graph}(d)]. 
The next term describes higher-order contributions, e.g., 
a hopping process $0 \to 1 \to 2 \to 0$ 
with propagations within each branch connected with the sites $1$ and $2$.  
The expansion (\ref{perturb}) is simplified and reduced to a set of recursion equations
owing to the self-similarity of THC under the ice rule: $g_i$ depends not explicitly on $i$ but only on the value of $U_i$; i.e., $g_i = g_{\pm}$ corresponding to $U_i = \pm U/2$. The recursive equations are given by
\begin{eqnarray}
\frac{g_\pm(1-g_\mp)+2g_\mp(1-g_\pm)}{1+g_\mp(1-2g_\pm)} 
= \frac{1}{t}\big( \varepsilon \mp \frac{U}{2} \big) - \frac{1}{g_\pm}.
\label{g_Dyson}
\end{eqnarray}
The full Green's functions $G_{\pm}$ are similarly obtained as
\begin{eqnarray}
G_\pm^{-1} = \frac{2}{g_\pm} - \frac{1}{t}\big(\varepsilon \mp \frac{U}{2} \big).
\label{fullDyson}
\end{eqnarray}
Equations~(\ref{g_Dyson}) and (\ref{fullDyson}) give the exact solution to 
the local Green's functions of the extended Falicov-Kimball model (\ref{Ham}) in the ice-rule limit: $\sum_i n^f_i = N/2$ and $V \to \infty$.
By extending the above calculations, it is also possible to obtain the nonlocal as well as the finite-$T$ Green's functions.
The whole procedure can be straightforwardly applied to a broader range of models 
with general ice-rule type constraints on general cacti of a complete graph. 
Such extensions will be discussed elsewhere. 

\begin{figure}[b]
\begin{center}
\includegraphics[width=0.4\textwidth]{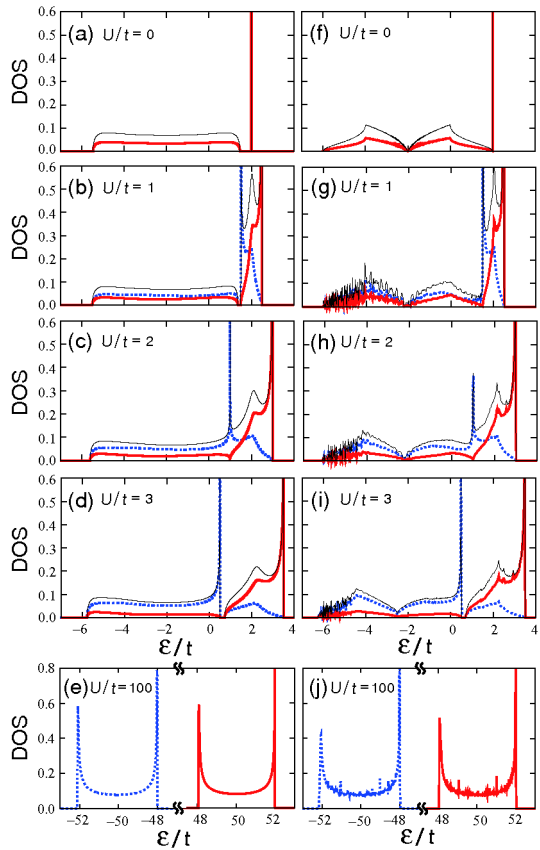}
\end{center}
\caption{\label{DOSplot} 
(color online). 
DOS for (a)-(e) THC and (f)-(j) pyrochlore lattice at $U/t=0$, 1, 2, 3, and 100. 
The results for the pyrochlore model are calculated for 
a $3^3$ superlattice of $4 \times 8^3$ lattice sites~\cite{note-ice}. 
$\rho_+(\varepsilon)$ [$\rho_-(\varepsilon)$] is shown by
the bold red (dotted blue) curves. 
In (a)-(d) and (f)-(i), the total DOS $\rho(\varepsilon)$ is plotted by thin black curves. 
}
\end{figure}

Now let us discuss the ground-state properties derived from the exact solution. 
Figure~\ref{DOSplot} shows the site-resolved density of states (DOS) 
given by $\rho_{\pm} = -{\rm Im} G_{\pm}/\pi t$ 
and their summation $\rho = \rho_+ + \rho_-$.
In Fig.~\ref{DOSplot}(a), we show DOS at $U=0$, 
which is given by the analytic form of 
$
\rho(\varepsilon) = \frac{1}{\pi}{\rm Re}\{[\sqrt{12t^2-(\varepsilon+2t)^2}]/[16t^2-(\varepsilon+2t)^2]\} + \frac{1}{2}\delta(\varepsilon-2t).
$
This is composed of the delta-functional peak (flat bands) at $\varepsilon= 2t$ and 
the broad spectrum for $|\varepsilon + 2t| < 2\sqrt{3}t$. 
In the opposite limit of $U \gg t$ also, a simple analytic form is available: 
By approximating $g_{\pm} \simeq \pm 0$ 
for $\varepsilon\simeq\mp U/2$ in Eq.~(\ref{g_Dyson}), we obtain
$
\rho_{\pm}(\varepsilon) = \frac{1}{\pi} 
{\rm Re} \{ 4t^2- [\varepsilon\mp(U/2)]^2 \}^{-1/2}. 
$
This is identical to DOS for one-dimensional (1D) tight-binding chains 
centered at $\varepsilon=\pm U/2$. The coincidence comes from the fact that 
THC under the ice-rule constraint is broken up into 1D chains of the same potential $+U/2$ or $-U/2$, 
as schematically shown in Fig.~\ref{graph}(b). 
A typical result is shown in Fig.~\ref{DOSplot}(e) at $U=100t$.

In both the limiting cases of $U=0$ and $U \gg t$, 
the system is insulating at half filling of the spinless fermion, $n^c \equiv \langle n^c_i \rangle = 1/2$. 
However, the origins of the two insulating states are quite different. The insulating phase at $U=0$ 
is a simple band insulator, in which the broad spectrum is fully occupied and the flat band is empty. 
On the other hand, the large-$U$ phase is an incompressible state with a large gap $\sim U$ 
originating from the repulsive interaction between fermions and immobile particles. We 
call this correlation-driven insulating state ``charge ice", in analogy with the spin ice~\cite{Harris,Ramirez}, 
since the mobile fermions are localized in the ice-rule configuration which is composed of the sites with $n_i^f=0$. 

\begin{figure}[b]
\begin{center}
\includegraphics[width=0.43\textwidth]{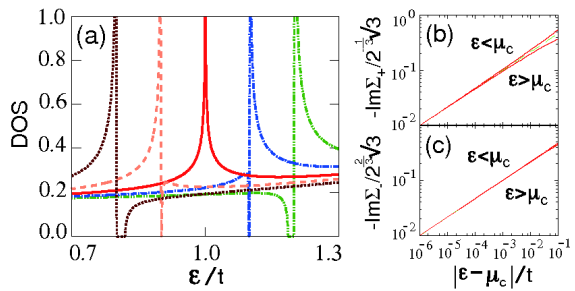}
\end{center}
\caption{\label{sigma} 
(color online). 
(a) DOS around $\varepsilon=\mu_c=t$ 
at $U/t=1.6$ (dash-double-dotted), $1.8$ (dash-dotted), $2.0$ (solid), $2.2$ (dashed), and $2.4$ (dotted).
(b) and (c) Non-Fermi-liquid behaviors 
of Im$\Sigma_\pm$  
[Eq.~(\ref{singularity})].  
The green dashed lines show the asymptotic power law 
$\propto |\varepsilon-\mu_c|^{\frac13}$.
}
\end{figure}

The question is how the system changes from the band insulator to the ``charge ice" as $U$ increases. 
In Figs.~\ref{DOSplot}(b)-(d), we show the change of DOS obtained in the intermediate range of $U$. 
When switching on $U$, while the broad spectrum is almost unchanged, the flat bands 
are perturbed to be broadened around $\varepsilon = 2t$, 
leading to a reduction of the gap [Fig.~\ref{DOSplot}(b)]. 
The gap decreases as $U$ increases and 
vanishes at $U \simeq 2t$ [Fig.~\ref{DOSplot}(c)]. 
Further increase of $U$ opens a gap again [Fig.~\ref{DOSplot}(d)], 
and increases it continuously. Critical behavior around $U=2t$ is shown in Fig.~\ref{sigma}(a). 
The plot suggests that the gap closes only at $U=U_c=2t$. 

\begin{figure}[t]
\begin{center}
\includegraphics[width=0.43\textwidth]{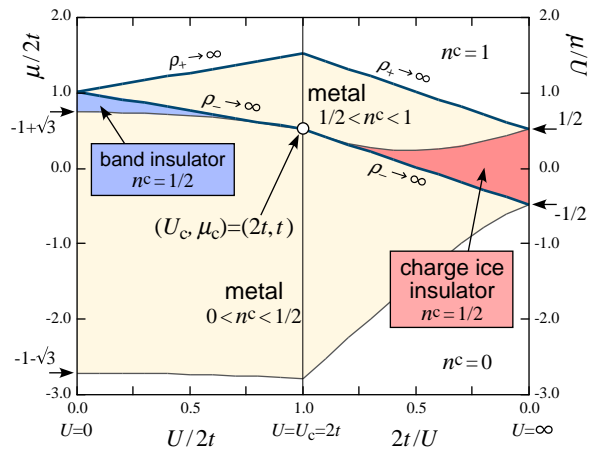}
\end{center}
\caption{\label{diagram} 
(color online). 
Ground-state phase diagram of 
the extended Falicov-Kimball model on THC in the ice-rule limit. 
}
\end{figure}

The critical behaviors of DOS can be understood analytically
from Eqs.~(\ref{g_Dyson}) and (\ref{fullDyson}).
We can prove that the divergence occurs at $\varepsilon = \varepsilon_U = t + (U_c - U)/2$: 
At $\varepsilon = \varepsilon_U$, 
we obtain 
$g_+^{-1} = -1 - (U-U_c)/t$, $g_-^{-1} = 1$, $G_+ = -t/U$, and $G_- = \infty$, resulting in 
$\rho_+(\varepsilon_U)=0$ (except for $U=0$) while $\rho_-(\varepsilon_U)\rightarrow\infty$.
With regard to the gap, by considering a small deviation from $U=U_c$ and 
evaluating $g_{\pm}$ at $\varepsilon \simeq \varepsilon_U$, we find that the energy gap opens as 
$\Delta(U) \simeq \frac{8}{27t^2}|U-U_c|^3$ for both $U > U_c$ and $U < U_c$. Therefore the energy gap closes only at $U=U_c$. 
This is identified as QCP between the band insulator and the charge ice (see the phase diagram in Fig.~\ref{diagram}).

The critical behavior of the gap $\Delta(U) \propto |U-U_c|^3$ is peculiar 
in contrast to the usual linear behavior $\Delta(U) \propto |U-U_c|$ 
in the Mott transition~\cite{Imada1994}. 
Actually, QCP is peculiar also in the sense that the self-energy exhibits an anomalous power-law behavior. 
From Eqs.~(\ref{g_Dyson}) and (\ref{fullDyson}), 
we can derive that the self-energy $\Sigma_{\pm}$, defined by 
$\Sigma_{\pm} = g_{\pm}^{(0)-1} - G_{\pm}^{-1}$, shows the following critical behavior:
\begin{eqnarray}
{\rm Re}\Sigma_{\pm}(\varepsilon) &=& 2
- C_{\pm}|\varepsilon-\mu_c|^{\frac{1}{3}}{\rm{sgn}}(\varepsilon-\mu_c),\\
{\rm Im}\Sigma_{\pm}(\varepsilon) &=& -C_{\pm}\sqrt{3} \, |\varepsilon-\mu_c|^{\frac{1}{3}},
\label{singularity}
\end{eqnarray}
where $C_-=(4/t)^{\frac{1}{3}}$, $C_+ = C_-/2$, and 
$\mu_c = t$ is the critical chemical potential [Figs.~\ref{sigma}(b) and (c)]. 
The anomalous power law $\propto |\varepsilon-\mu_c|^{\frac{1}{3}}$
indicates that the system shows a non-Fermi-liquid behavior at QCP. DOS also shows a singular energy dependence, 
$\rho_{\pm}(\varepsilon)\propto |\varepsilon-\mu_c|^{\pm\frac{1}{3}}$ 
at $U=U_c$, resulting in the anomalous $T$ dependence of thermodynamic quantities at QCP. 
For example, the specific heat is predicted to behave as $\propto T^{\frac23}$ at low $T$.

Collecting the results with varying $U$, 
we summarize the exact ground-state phase diagram in Fig.~\ref{diagram}. 
There are two metallic regions for $0 < n^c < 1/2$ and 
$1/2 < n^c < 1$, 
which are separated by the two insulating regions at half filling $n^c = 1/2$, 
i.e., the band insulator for $U<U_c$ and the charge ice for $U>U_c$. 
All these four phases meet at QCP at $(U_c, \mu_c) = (2t, t)$. 

As indicated by the bold curve crossing QCP in Fig.~\ref{diagram}, 
the divergence of $\rho_-$ at $\varepsilon = \varepsilon_U$ is transferred from the upper-band bottom to the lower-band top 
[see Fig.~\ref{sigma}(a)]. The divergence contains a bunch of extended states on the sites with $n_i^f=0$. 
The extended states are fully occupied in the charge ice state, 
while they are empty in the band insulator.  
This fact leads us to define an ``order parameter" 
to distinguish two insulating states, 
$O_{\Psi}\equiv\langle c^{\dag}_{\Psi}c_{\Psi}\rangle$ for the corresponding extended-state operator 
$
c_{\Psi} 
= \sum_i c_i (-1)^i / \sqrt{L}, 
$
where $i$ represents the sequential site number on a chain composed of $n_i^f=0$ sites 
($L$ is the length). $O_{\Psi}$ changes from 0 for the band insulator  
to 1 for the charge ice. The transfer of the extended states bears some analogy to the 
``levitation scenario" proposed for the quantum Hall systems~\cite{Laughlin}.
This analogy suggests a discrete change of the ``transport nature" at QCP. 
Further analysis will be discussed elsewhere.

Finally, we return to consider the original pyrochlore model. As indicated in Fig.~\ref{DOSplot}, 
we observe many similar behaviors in DOS between THC and pyrochlore models~\cite{note-ice,Ishizuka}:   
(i) DOS consists of the flat bands and dispersive bands at $U=0$, 
(ii) $U$ drives a quantum phase transition to the charge ice insulator at half filling, 
(iii) the divergence of DOS transfers through QCP, and 
(iv) DOS shows a one-dimensional-like form for $U \gg t$. 
These are direct consequences from the key features shared between THC and pyrochlore, 
i.e., the corner-sharing network of tetrahedra and the resulting macroscopic ice-rule degeneracy.
Interestingly enough, (iii) suggests a possibility that the transition in the pyrochlore case is also described 
by an order parameter analogous to $O_{\Psi}$, accompanied by similar anomalous critical behavior.
We note that, in the weak $U$ region, THC is a band insulator, whereas the pyrochlore model is metallic at half filling [Figs.~\ref{DOSplot}(f)-(h)]; 
however, this difference is rather irrelevant since our focus is on the strongly correlated physics 
related to the charge ice insulator. The benefit from obtaining the exact solution exceeds the minor dissimilarity. 
Thus, our THC solution captures the common essential physics of the itinerant ice-rule systems. 
Further comparisons, including the effect of global loops neglected in THC, will be reported separately~\cite{Ishizuka}. 

In summary, we have exactly solved the extended Falicov-Kimball model on the tetrahedron Husimi cactus in the ice-rule limit. 
The solution reveals a quantum critical point and associated non-Fermi-liquid behavior
in quantum melting of ``charge ice" insulator.
Furthermore, the results capture many essential features of more realistic lattices 
with corner-sharing tetrahedra, such as the pyrochlore lattice. 
Our exact solution provides a canonical reference to the itinerant ice-rule physics, 
and will open new avenues of research with wide applicability to 
correlation-induced phenomena under strong frustration. 

The authors thank I.\ Maruyama for fruitful discussions.
This work was supported by KAKENHI (Nos. 17071003, 19052008, 21740242, and 21340090), 
the Global COE Program ``the Physical Sciences Frontier", 
and by the Next Generation Super Computing Project, Nanoscience Program, MEXT, Japan.


\begin{thebibliography}{9}
\bibitem{note-ice} 
In the pyrochlore model, the ice degeneracy will be ultimately lifted in the ground state.
Because of the lack of the exact ground state, we here simply calculate DOS
for Eq. (2) with taking average over 40 different ice configurations randomly chosen.
The results are well converged and supposed to represent the situation at low but finite $T$
higher than the energy scale of the degeneracy lifting, which turns out 
to be remarkably small $\sim 10^{-4}-10^{-3}t$ in the entire range of $U$~\cite{Ishizuka}.
\end{thebibliography}
\end{document}